\documentclass[pre,aps,amsmath,amssymb,reprint,superscriptaddress]{revtex4-2}

\usepackage{graphicx}
\usepackage{dcolumn}
\usepackage{bm}
\usepackage[
    bookmarks=false,
    colorlinks=true,
    linkcolor=blue,
    citecolor=blue,
    urlcolor=blue
]{hyperref}
\usepackage{orcidlink}

\definecolor{ppt-blue}{RGB}{2, 83, 118}

\begin{document}

\title{
%
    Statistical mechanical model for crack growth
}
\author{
    Michael R. Buche%
    \:\orcidlink{0000-0003-1892-0502}\,
}
\email{mrbuche@sandia.gov}
\affiliation{
    Computational Solid Mechanics and Structural Dynamics, Sandia National Laboratories, Albuquerque, New Mexico 87185, USA
}
\author{
    Scott J. Grutzik%
    \:\orcidlink{0000-0002-6490-3941}\,
}
\affiliation{
    Materials and Failure Modeling, Sandia National Laboratories, Albuquerque, New Mexico 87185, USA
}
\date{
    \today
}

\begin{abstract}
Analytic relations that describe crack growth are vital for modeling experiments and building a theoretical understanding of fracture.
Upon constructing an idealized model system for the crack and applying the principles of statistical thermodynamics, it is possible to formulate the rate of thermally activated crack growth as a function of load, but the result is analytically intractable.
Here, an asymptotically correct theory is used to obtain analytic approximations of the crack growth rate from the fundamental theoretical formulation.
These crack growth rate relations are compared to those that exist in the literature and are validated with respect to Monte Carlo calculations and experiments.
The success of this approach is encouraging for future modeling endeavors that might consider more complicated fracture mechanisms, such as inhomogeneity or a reactive environment.
\smallskip\smallskip\smallskip

\noindent DOI: \href{https://doi.org/10.1103/PhysRevE.109.015001}{10.1103/PhysRevE.109.015001}
\end{abstract}

\maketitle


\section{Introduction}\label{sec:introduction}

Fracture is a direct result of breaking atomic bonds, and it is therefore critical to include microscopic physics in macroscopic models for crack growth.
Though the fundamental theory of fracture mechanics, formulated using continuum thermodynamics \cite{griffith1921vi}, is quite successful, the theory cannot explain why fracture occurs by relating it to atomic properties \cite{lawn1983physics}.
This shortcoming prevents continuum fracture mechanics models from addressing the significant impacts of the discrete microstructure \cite{thomson1971lattice,sinczair1972atomistic,sinclair1972atomistic,sinclair1975influence}, thermal energy and kinetic effects \cite{brenner1962mechanical,zhurkov1965kinetic,lawn1975atomistic,cook1993kinetics}, chemical interaction \cite{fuller1980atomic,wiederhorn1980micromechanisms,michalske1983molecular}, or unstable dynamic propagation \cite{marder1993simple,marder1993instability,marder1995origin,gorbushin2019dynamic}.

To accurately model and investigate these atomistic mechanisms, a substantial amount of work has been accomplished over the past half century using both analytic models \cite{marder2015particle} and fully atomistic simulations \cite{bitzek2015atomistic}.
Starting with the model of \citet{thomson1971lattice}, several existing models use a quasi-one-dimensional discrete arrangement of particles to represent crack faces, which are then treated mechanically \cite{thomson1971lattice,sinczair1972atomistic,sinclair1972atomistic,sinclair1975influence}.
Since thermal energy and the related kinetic effects are important, especially in the subcritical regime \cite{santucci2003thermal,santucci2007slow,vanel2009time}, these and other models have been augmented by assuming an Arrhenius \cite{arrhenius1889reaktionsgeschwindigkeit}, Eyring \cite{eyring1935activated}, or Kramers \cite{kramers1940brownian} rate \cite{sinclair1975influence,fuller1980atomic,wiederhorn1980micromechanisms,lawn1983physics, cook1993kinetics,cook2019thermal,krausz1988fracture,lawn1993fracture,michalske1983molecular,ciccotti2009stress,le2009subcritical,grutzik2022kinetic}.
Similar models have been applied to interfaces \cite{maddalena2009mechanics,wei2014stochastic,qian2017thermally,yang2019rate,yang2020multiscale,lei2022multiscale}.
\linebreak
To properly include temperature effects and the associated kinetics, it would be better to incorporate statistical thermodynamics in the model from the start.
Several models have used statistical physics \cite{marder1995fluctuations,marder1996statistical,marder2004effects,freund2009characterizing,freund2014brittle}, some even modeling fracture as a phase transition \cite{selinger1991statistical,buchel1997statistical,alava2006statistical}, but they do not compute the partition function necessary for statistical thermodynamics.
Simulations like molecular dynamics use a complete atomistic description of fracture \cite{sinclair1978flexible,bernstein2003lattice,luo2021atomic,buze2021numerical} and can even explicitly model chemical reactions \cite{rimsza2018chemical,rimsza2022inelastic,rimsza2022water} or run concurrently in a multiscale approach \cite{gu2021dissolution,zhao2022atomic}, but they lack analytic interpretability.

Although considerable progress has been made in the atomic scale modeling of crack growth, further progress is necessary in the area of analytic model development.
While the principles of statistical thermodynamics have been utilized to analytically model crack growth, they have not been rigorously applied to the atomistic model systems that were only treated mechanically \cite{thomson1971lattice,sinczair1972atomistic,sinclair1972atomistic,sinclair1975influence,fuller1980atomic,lawn1983physics}.
Such a treatment, as accomplished here, is vital for incorporating the simultaneity of both mechanical and thermal effects, especially in subcritical crack growth regimes.
Beginning from the definition of the crack model system via the Hamiltonian, the partition function is formulated.
Analytic relations are obtained for the rate of crack growth which are asymptotically valid for steep bonded potentials ahead of the crack tip \cite{buche2021fundamental,buche2022freely,buche2023modeling}, and are numerically verified using Monte Carlo calculations.
Both the isometric and isotensional thermodynamic ensembles are considered, and the thermodynamic limit of large system size.
This model and its applications are distinctly different from cohesive zone models \cite{barenblatt1959concerning,dugdale1960yielding,sharpe2008springer,liu2013cohesive}, and unlike the model of \citet{marder1995fluctuations,marder1996statistical}, it does not support steady state crack growth.
Ultimately, an asymptotic relation for the subcritical crack growth velocity is obtained,

\begin{equation}\label{eq:knet}
    v \sim \frac{b\omega_0}{\pi}\, \exp\left(\frac{f\Delta x^\ddagger - \Delta u^\ddagger}{kT}\right) \sinh\left(\frac{Rb^2}{2kT}\right)
    ,
\end{equation}
where $b$ is the atom spacing, $\omega_0$ is the attempt frequency, $\Delta u^\ddagger$ is the potential energy barrier to the transition state for breaking a bond, $R$ is the energy release rate, $k$ is the Boltzmann constant, and $T$ is the temperature.
The key difference here between Eq.~\eqref{eq:knet} and many past relations \cite{lawn1975atomistic,wiederhorn1980micromechanisms,cook1993kinetics,krausz1988fracture,michalske1983molecular,ciccotti2009stress,le2009subcritical,cook2019thermal,grutzik2022kinetic} is an emergent \citet{bell1978models} term $e^{f\Delta x^\ddagger}$ similar to \citet{marder1996statistical}, where $f\equiv\sqrt{REb^3}$ is the force, $E$ is the elastic modulus, and $\Delta x^\ddagger$ is the transition state bond displacement.
Eq.~\eqref{eq:knet} is verified numerically and with respect to subcritical crack growth experiments.

\clearpage

\section{Model system}\label{sec:model}

\begin{figure}[t]
    \begin{center}
        \includegraphics{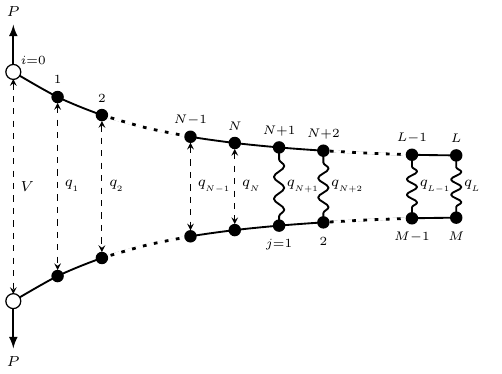}
    \end{center}
    \caption{\label{fig:model-system}%
        The crack model system.
        The statistical mechanical ensemble is characterized by the number of repeat units behind the crack tip $N$, the number ahead of and including the crack tip $M$, either the (massless) end separation $V$$\equiv$\,$q_0$ or force $P$, and the temperature $T$.
        With symmetry, the $2L$ system degrees of freedom ($L$=$N$+$M$) are the $L$ crack face separations $q_i$ and the $L$ corresponding momenta $p_i$.
    }
\end{figure}

The crack is represented by a discrete set of particles, connected along the crack faces by bending elements and, ahead of the crack tip, connected across the crack plane by bond elements \cite{thomson1971lattice,fuller1980atomic}.
On either crack face, let there be $N$ particles behind the crack tip, and $M$ particles ahead of and including the crack tip, for $L=N+M$ total particle pairs; see Fig.~\ref{fig:model-system}.
Assuming that the system remains symmetric about the crack plane, the relevant degrees of freedom are the crack face separations $q_i$ and the corresponding momenta $p_i=m\dot{q}_i$, where $m$ is the reduced mass.
The Hamiltonian of the system is then

\begin{equation}
    H(\mathbf{p}, \mathbf{q}) =
    \sum_{i = 1}^{L} \frac{p_i^2}{2m} + U(\mathbf{q})
    ,
\end{equation}
where $U(\mathbf{q}) = U_0(\mathbf{q}) + U_1(\mathbf{q})$ is the system potential energy.
The system potential energy from bending is

\begin{equation}\label{eq:U_0}
    U_0(\mathbf{q}) =
    \sum_{i = 2}^{L}
        \frac{c}{2} \left(q_{i - 2} - 2q_{i - 1} + q_i\right)^2
    ,
\end{equation}
where $c$ is the bending element stiffness.
This term can be thought of as representing the coupling to a linear elastic bulk \cite{fuller1980atomic}.
Note that $V\equiv q_0$ is fixed when the end separation $V$ is prescribed (isometric), but not when the end force $P$ is prescribed (isotensional).
The system potential energy from stretching bonds is

\begin{equation}\label{eq:U_1}
    U_1(\mathbf{q}) =
    \sum_{i = N + 1}^L u(q_i) =
    \sum_{j = 1}^M u(q_j)
    ,
\end{equation}
where $u(q)$ is the potential energy function for a bond.
Here the Morse potential \cite{morse1929diatomic} is used, given by

\begin{equation}\label{eq:u}
    u(q) = 
    u_0 \left[ 1 - e^{-a(q - b)} \right]^2
    ,
\end{equation}
where $u_0$ is the bond energy, $b$ is the equilibrium bond length, and $a$ is the Morse parameter.
Note that this system resembles some describing other physical phenomena, notably the unzipping of macromolecules \cite{peyrard1989statistical,dauxois1993dynamics,theodorakopoulos2004nonlinear,peyrard2004nonlinear,singh2005statistical,rapti2011transfer,da2019equilibrium,florio2020role,cannizzo2021temperature,florio2023predictive,cannizzo2023thermal,bellino2023cooperative}.

\subsection{Isometric ensemble}\label{sec:model.isometric}

The isometric ensemble partition function is given by

\begin{equation}\label{eq:Q}
    Q(N,M,V,T) =
    \frac{1}{h^L} \int dp \int dq \ e^{-\beta H(\mathbf{p}, \mathbf{q})}
    ,
\end{equation}
where $h$ is the Planck constant and $\beta = 1/kT$, where $T$ is the temperature and $k$ is the Boltzmann constant \cite{mcq}.
Eq.~\eqref{eq:Q} can be decomposed as a product $Q=Q_\mathrm{mom}Q_\mathrm{con}$, where the momentum integral evaluates to

\begin{equation}\label{eq:Q_mom}
    Q_\mathrm{mom}(N,M,T) =
    \left(\frac{2\pi m}{\beta h^2}\right)^{L/2}
    ,
\end{equation}
and where the configuration integral is given by

\begin{equation}\label{eq:Q_con}
    Q_\mathrm{con}(N,M,V,T) =
    \int dq \ e^{-\beta U(\mathbf{q})}
    .
\end{equation}
Dependence on $(N,M,V,T)$ is implicit in the following.
The Helmholtz free energy $A$ and expected end force $P$ are, respectively, given by

\begin{equation}\label{eq:AP}
    A = -\frac{1}{\beta}\,\ln Q
    ,\qquad
    P = \frac{\partial A}{\partial V}
    .
\end{equation}
Applying transition state theory \cite{zwanzig2001nonequilibrium}, if $q^\ddagger$ is the transition state length of the crack tip bond, the rate of breaking the crack tip bond to advance the crack is given by

\begin{equation}\label{eq:k.isometric}
    k' =
    \sqrt{\frac{1}{2\pi m \beta}} \ \frac{Q_\mathrm{con}^\ddagger}{Q_\mathrm{con}}
    ,
\end{equation}
where the transition state configuration integral is

\begin{equation}\label{eq:Q_con_ts}
    Q_\mathrm{con}^\ddagger = 
    \int dq \ \delta\left(q_{N+1} - q^\ddagger\right) \, e^{-\beta U(\mathbf{q})} 
    .
\end{equation}
Here $\delta$ is the Dirac delta function, and the transition state location $q^\ddagger$ is chosen to correspond to the maximum force for the Morse potential \cite{buche2021chain}, which is $q^\ddagger=b+\ln(2)/a$.

\subsection{Isotensional ensemble}\label{sec:model.isotensional}

The isotensional ensemble partition function is given by $Z=Z_\mathrm{mom}Z_\mathrm{con}$, where $Z_\mathrm{mom}=Q_\mathrm{mom}$ and

\begin{equation}\label{eq:Z_con}
    Z_\mathrm{con}(N,M,P,T) =
    \int dV \int dq \ e^{-\beta\Pi(V,\mathbf{q})}
    ,
\end{equation}
where $\Pi=U-PV$ is the system total potential energy.
Dependence on $(N,M,P,T)$ is implicit in the following.
The Gibbs free energy $G$ and expected end position $V$ are respectively given by

\begin{equation}\label{eq:GV}
    G = -\frac{1}{\beta}\,\ln Z
    ,\qquad
    V = -\frac{\partial G}{\partial P}
    .
\end{equation}
Again applying transition state theory \cite{zwanzig2001nonequilibrium}, the rate of breaking the crack tip bond to advance the crack is

\begin{equation}\label{eq:k.isotensional}
    k' =
    \sqrt{\frac{1}{2\pi m \beta}} \ \frac{Z_\mathrm{con}^\ddagger}{Z_\mathrm{con}}
    ,
\end{equation}
where the transition state configuration integral is

\begin{equation}\label{eq:Z_con_ts}
    Z_\mathrm{con}^\ddagger = 
    \int dV \int dq \ \delta\left(q_{N+1} - q^\ddagger\right) \, e^{-\beta\Pi(V,\mathbf{q})} 
    .
\end{equation}

\subsection{Thermodynamic limit}\label{sec:model.thermodynamic}

Generally, results in either the isometric or isotensional ensembles will differ, such as the expected mechanical response or the rate of breaking the crack tip bond.
Upon referencing other systems \cite{mcq,neumann2003precise,suzen2009ensemble,manca2014equivalence,buche2020statistical}, it is reasonable to expect that these differences will vanish as the system becomes large.
For example, the Legendre transformation is likely valid for many repeat units both behind and ahead of the crack tip and appreciable loads,

\begin{equation}\label{eq:Legendre}
    G \sim A - PV
    \quad \text{for } N,M\gg 1
    .
\end{equation}
This limit of large system size, in which the results of either thermodynamic ensemble asymptotically become equivalent, is referred to as the thermodynamic limit.

\section{Asymptotic approach}\label{sec:asymptotic}

\begin{table}[b]
    \caption{\label{table:nondimensionalize}%
        Nondimensional variables for the crack model system presented in Sec.~\ref{sec:model}.
        The nondimensional transition state stretch and Morse parameter are related via $\lambda^\ddagger\equiv1+\ln(2)/\alpha$.
    }
    \begin{ruledtabular}
        \begin{tabular}{lcc}
            & Dimensional & Nondimensional \\
            \colrule
            Crack face separations & $q_i$ & $s_i\equiv q_i/b$ \\
            Bond lengths & $q_j$ & $\lambda_j\equiv s_{N+j}$ \\
            Bending stiffness & $c$ & $\kappa\equiv \beta cb^2$ \\
            Bond energy & $u_0$ & $\varepsilon\equiv\beta u_0$ \\
            Morse parameter & $a$ & $\alpha\equiv ab$ \, \\
            End separation & $V$ & $v\equiv V/b$ \\
            End force & $P$ & $p\equiv \beta Pb$ \\
        \end{tabular}
    \end{ruledtabular}
\end{table}

Since the configuration integrals in the previous section cannot be evaluated with any ease, accurate asymptotic approximations are now developed.
These asymptotic relations are entirely analytic and closed-form and therefore maintain both efficiency and interpretability in contrast to simulation or numerical integration approaches.
Essentially, the asymptotic approach approximates the statistical thermodynamics of the full system by building upon an analytically tractable reference system, where the approximation error vanishes as the relevant potentials become steep \cite{buche2021fundamental,buche2022freely,buche2023modeling}.
To begin, the variables for the crack model system are nondimensionalized in Table~\ref{table:nondimensionalize}.
Applying these nondimensional variables to Eq.~\eqref{eq:U_0}, the nondimensional potential energy from bending the crack faces is given by

\begin{equation}\label{eq:betaU_0}
    \beta U_0(\mathbf{s}) =
    \sum_{i = 2}^L
        \frac{\kappa}{2} \left(s_{i - 2} - 2s_{i - 1} + s_i\right)^2
    .
\end{equation}
Applying the set of nondimensional variables in Table~\ref{table:nondimensionalize} to Eqs.~\eqref{eq:U_1} and \eqref{eq:u}, the nondimensional potential energy from stretching bonds ahead of and including the crack tip is given by

\begin{equation}
    \beta U_1(\boldsymbol{\lambda}) =
    \sum_{j = 1}^M
        \varepsilon \left[ 1 - e^{-\alpha(\lambda_j - 1)} \right]^2
    .
\end{equation}
In the following subsections, it becomes useful to split Eq.~\eqref{eq:betaU_0} into two separate contributions, $U_0=U_{00}+U_{01}$.
Crucially, this split allows the separation of the system potential energy ahead of and including the crack tip ($U_{01}+U_1$) from the potential energy governing the reference system ($U_{00}$).
Note that $U_{00}$ is the potential energy function for the discrete representation of a linear elastic slender beam in bending, where fixing $s_{N+1}\equiv\lambda_1$ and $s_{N+2}\equiv\lambda_2$ would then specify the boundary conditions.
The contribution from bending the crack faces behind and including the crack tip is

\begin{equation}\label{eq:betaU_00}
    \beta U_{00}(\mathbf{s}) \equiv
    \sum_{i = 0}^N
        \frac{\kappa}{2} \left(s_{i} - 2s_{i + 1} + s_{i + 2}\right)^2
    ,
\end{equation}
and the contribution from bending the crack faces ahead of the crack tip is

\begin{equation}\label{eq:betaU_01}
    \beta U_{01}(\boldsymbol{\lambda}) \equiv
    \sum_{j = 3}^M
        \frac{\kappa}{2} \left(\lambda_{j - 2} - 2\lambda_{j - 1} + \lambda_j\right)^2
    .
\end{equation}

\subsection{Isometric ensemble}\label{sec:asymptotic.isometric}

The isometric configuration integral for the full system in Eq.~\eqref{eq:Q_con} can be rewritten as

\begin{equation}\label{eq:Q_con_rewrite}
    Q_\mathrm{con}(v) =
    \int d\lambda \ Q_\mathrm{0,con}(v,\boldsymbol{\lambda}) \, e^{-\beta U_1(\boldsymbol{\lambda})}
    ,
\end{equation}
where the configuration integral for the reference system is, using Eqs.~\eqref{eq:betaU_00}--\eqref{eq:betaU_01}, defined as

\begin{equation}\label{eq:Q_0_con}
    Q_\mathrm{0,con}(v,\boldsymbol{\lambda}) \equiv
    e^{-\beta U_{01}(\boldsymbol{\lambda})} \int ds_1\cdots ds_N \ e^{-\beta U_{00}(\mathbf{s})}
    .
\end{equation}
The reference system here is the statistical mechanical treatment of the discrete representation of a linear elastic slender beam with a fixed end displacement $\Delta v\equiv v-1$.
Choosing the fixed bond stretches $\boldsymbol{\lambda}$ effectively specifies the boundary conditions (via $\lambda_1$ and $\lambda_2$) and translates the potential energy level.
As shown in Appendix~\ref{sec:ref-sys.isometric}, the integrals in Eq.~\eqref{eq:Q_0_con} can be computed analytically.
The result is

\begin{equation}\label{eq:Q_0_con_result}
    Q_\mathrm{0,con}(v,\boldsymbol{\lambda}) =
    \sqrt{\frac{(2\pi)^N}{\det\mathbf{H}}} \ e^{\tfrac{1}{2}\mathbf{g}^T\cdot\mathbf{H}^{-1}\cdot\mathbf{g} - f - \beta U_{01}}
    ,
\end{equation}
where $\mathbf{H}$ is the Hessian of $\beta U_{00}$ with respect to the set of variables $\{s_1,\ldots,s_N\}$, which has the components

\begin{equation}\label{eq:H.isometric}
    H_{mn} =
    \kappa\left(
        6\delta^m_n - \delta^m_1\delta^n_1
        - 4\delta^{|m-n|}_1 + \delta^{|m-n|}_2
    \right)
    ,
\end{equation}
and where $\mathbf{g}$ and $f$ are given by

\begin{equation}\label{eq:gf.isometric}
    \begin{aligned}
        \mathbf{g}(v,\lambda_1,\lambda_2) =& \
            \kappa\left(
                2v,-v,0,\ldots,0,-\lambda_1,4\lambda_1-\lambda_2
            \right)^T
        ,\\
        f(v,\lambda_1,\lambda_2) =& \
            \frac{\kappa}{2}\left[
                v^2 + \lambda_1^2 + \left(2\lambda_1 - \lambda_2\right)^2
            \right]
        .
    \end{aligned}
\end{equation}
The reference system configuration integral $Q_\mathrm{0,con}(v,\boldsymbol{\lambda})$ is now known analytically via Eqs.~\eqref{eq:Q_0_con_result}--\eqref{eq:gf.isometric}, but the full system configuration integral $Q_\mathrm{con}(v)$ in Eq.~\eqref{eq:Q_con_rewrite} still cannot be computed analytically.
Alternatively, the asymptotic approach of \citet{buche2021fundamental} is readily applicable to the form of $Q_\mathrm{con}(v)$ in Eq.~\eqref{eq:Q_con_rewrite} and will produce an analytic approximation \cite{buche2021fundamental,buche2022freely,buche2023modeling}.
Assuming that the bonded potentials $u$ constituting $U_1$ are all steep ($\varepsilon\gg 1$), the configuration integral for the full system may be asymptotically related to that for the reference system as

\begin{equation}\label{eq:Q_con_asymp}
    Q_\mathrm{con}(v) \sim
    Q_\mathrm{0,con}(v,\hat{\boldsymbol{\lambda}}) \prod_{j=1}^M
        \sqrt{\frac{2\pi}{\beta u''(\hat{\lambda}_j)}} \ e^{-\beta u(\hat{\lambda}_j)}
    ,
\end{equation}
where the bond stretches $\hat{\boldsymbol{\lambda}}$ are from minimizing $\beta U$ with respect to $\{s_1,\ldots,s_L\}$.
To approximate the full system ($Q_\mathrm{con}$), the asymptotic approach essentially combines the results of the analytically solvable reference system ($Q_\mathrm{0,con}$) and the mechanical treatment of the full system (minimizing $\beta U$).
The athermal rigid constraints ($\boldsymbol{\lambda}$) of the reference system are replaced by asymptotically correct approximations for thermal fluctuations encountered by the steep potentials ($\varepsilon\gg 1$) in the full system, provided by Laplace's method \cite{bleistein1975asymptotic,bender2013advanced} about the potential energy minimum ($\hat{\boldsymbol{\lambda}}$).
In the limit that the relevant potentials become infinitely steep ($\varepsilon\to\infty$), these thermal fluctuations become negligible and the full system behaves as the reference system.

If $\Delta A\equiv A(v)-A(1)$, Eq.~\eqref{eq:AP} is nondimensionalized as

\begin{equation}\label{eq:betaDeltaAandp}
    \beta\Delta A(v) = \ln\left[\frac{Q_\mathrm{con}(1)}{Q_\mathrm{con}(v)}\right]
    ,\qquad
    p(v) = \frac{\partial\beta\Delta A}{\partial v}
    ,
\end{equation}
which become asymptotic approximations when Eq.~\eqref{eq:Q_con_asymp} is utilized.
Eq.~\eqref{eq:k.isometric} can similarly be used to obtain an asymptotic relation for $k'$, where the asymptotic relation for $Q_\mathrm{con}^\ddagger$ takes the same form as Eq.~\eqref{eq:Q_con_asymp}, with the following changes.
$\beta U$ must be minimized with $\lambda_1$ fixed at $\lambda^\ddagger$, generally resulting in a different $\hat{\boldsymbol{\lambda}}$, and the $j=1$ frequency term in the product must then be removed.

\subsection{Isotensional ensemble}\label{sec:asymptotic.isotensional}

The isotensional configuration integral for the full system in Eq.~\eqref{eq:Z_con} can be rewritten as

\begin{equation}\label{eq:Z_con_rewrite}
    Z_\mathrm{con}(p) =
    \int d\lambda \ Z_\mathrm{0,con}(p,\boldsymbol{\lambda}) \, e^{-\beta U_1(\boldsymbol{\lambda})}
    ,
\end{equation}
where the configuration integral for the reference system is, using Eqs.~\eqref{eq:betaU_00}--\eqref{eq:betaU_01} and $\beta\Pi_{00}\equiv \beta U_{00} - ps_0$, is

\begin{equation}\label{eq:Z_0_con}
    Z_\mathrm{0,con}(p,\boldsymbol{\lambda}) \equiv
    e^{-\beta U_{01}(\boldsymbol{\lambda})} \int ds_0\cdots ds_N \ e^{-\beta\Pi_{00}(\mathbf{s})}
    .
\end{equation}
The reference system here is the statistical mechanical treatment of the discrete representation of a linear elastic slender beam with a fixed end force $p$.
Choosing the fixed bond stretches $\boldsymbol{\lambda}$ effectively specifies the boundary conditions (via $\lambda_1$ and $\lambda_2$) and translates the total potential energy level.
As shown in Appendix~\ref{sec:ref-sys.isotensional}, the integrals in Eq.~\eqref{eq:Z_0_con} can be computed analytically.
The result is

\begin{equation}\label{eq:Z_0_con_result}
    Z_\mathrm{0,con}(p,\boldsymbol{\lambda}) =
    \sqrt{\frac{(2\pi)^{N+1}}{\det\mathbf{H}}} \ e^{\tfrac{1}{2}\mathbf{g}^T\cdot\mathbf{H}^{-1}\cdot\mathbf{g} - f - \beta U_{01}}
    ,
\end{equation}
where $\mathbf{H}$ is the Hessian of $\beta\Pi_{00}$ with respect to the set of variables $\{s_0,\ldots,s_N\}$, which has the components

\begin{equation}\label{eq:H.isotensional}
    \begin{aligned}
        H_{mn} =
        \kappa\left(
            6\delta^m_n - 5\delta^m_1\delta^n_1 - \delta^m_2\delta^n_2
            - 4\delta^{|m-n|}_1
        \right. & \\ \left.
            + 2\delta^m_1\delta^n_2 + 2\delta^m_2\delta^n_1 + \delta^{|m-n|}_2
        \right) &
        ,
    \end{aligned}
\end{equation}
and where $\mathbf{g}$ and $f$ are given by

\begin{equation}\label{eq:gf.isotensional}
    \begin{aligned}
        \mathbf{g}(p,\lambda_1,\lambda_2) =& \
            \kappa\left(
                p/\kappa,0,\ldots,0,-\lambda_1,4\lambda_1-\lambda_2
            \right)^T
        ,\\
        f(\lambda_1,\lambda_2) =& \
            \frac{\kappa}{2}\left[
                \lambda_1^2 + \left(2\lambda_1 - \lambda_2\right)^2
            \right]
        .
    \end{aligned}
\end{equation}
Assuming that the bonded potentials $u$ constituting $U_1$ are all steep ($\varepsilon\gg 1$), the configuration integral for the full system may be asymptotically related to that for the reference system as

\begin{equation}\label{eq:Z_con_asymp}
    Z_\mathrm{con}(p) \sim
    Z_\mathrm{0,con}(p,\hat{\boldsymbol{\lambda}}) \prod_{j=1}^M
        \sqrt{\frac{2\pi}{\beta u''(\hat{\lambda}_j)}} \ e^{-\beta u(\hat{\lambda}_j)}
    ,
\end{equation}
where the bond stretches $\hat{\boldsymbol{\lambda}}$ are from minimizing $\beta\Pi$ with respect to $\{s_0,\ldots,s_L\}$.
If $\Delta G\equiv G(p)-G(0)$, Eq.~\eqref{eq:GV} is nondimensionalized as

\begin{equation}\label{eq:betaDeltaGandv}
    \beta\Delta G(p) = \ln\left[\frac{Z_\mathrm{con}(0)}{Z_\mathrm{con}(v)}\right]
    ,\qquad
    v(p) = -\frac{\partial\beta\Delta G}{\partial p}
    ,
\end{equation}
which become asymptotic approximations when Eq.~\eqref{eq:Z_con_asymp} is utilized.
Eq.~\eqref{eq:k.isotensional} can similarly be used to obtain an asymptotic relation for $k'$, where the asymptotic relation for $Z_\mathrm{con}^\ddagger$ takes the same form as Eq.~\eqref{eq:Z_con_asymp}, with the follwing changes.
$\beta\Pi$ must be minimized with $\lambda_1$ fixed at $\lambda^\ddagger$, generally resulting in a different $\hat{\boldsymbol{\lambda}}$, and the $j=1$ frequency term in the product must then be removed.

\subsection{Thermodynamic limit}\label{sec:asymptotic.thermodynamic}

In the thermodynamic limit of large system size, the results of the asymptotic approach applied to either thermodynamic ensemble should be asymptotically equivalent.
To be clear, two asymptotic limits are considered: steep potentials ($\varepsilon\gg 1$) and a large system ($N,M\gg 1$).
For any finite (albeit large) system size, it is true that increasing applied loads can eventually cause large-system approximations to become inaccurate \cite{buche2020statistical}.
Therefore, the thermodynamic limit of large system size considered here also includes the notion of comparably small displacements or forces.
For example, the nondimensional displacement $\Delta v\equiv v-1$ applied in the isometric ensemble must be small compared to the nondimensional length $N$ in order for the large-system approximation of the mechanical response $p(v)$ to be accurate.

Applying the thermodynamic limit ($N,M\gg 1$) to the asymptotic relations ($\varepsilon\gg 1$) obtained in Sec.~\ref{sec:asymptotic.isometric} for the isometric ensemble, as shown in Appendix~\ref{sec:thermo-limit-asymp.isometric},

\begin{equation}\label{eq:asymptotic.AP}
    \beta\Delta A(v) \sim
    \frac{3\kappa}{2N^3}\left(\Delta v\right)^2
    ,\qquad
    p(v) \sim
    \frac{3\kappa}{N^3}\,\Delta v
    .
\end{equation}
Applying the same limit to the relations in Sec.~\ref{sec:asymptotic.isotensional} for the isotensional ensemble, as shown in Appendix~\ref{sec:thermo-limit-asymp.isotensional},

\begin{equation}\label{eq:asymptotic.GV}
    \beta\Delta G(p) \sim
    -\frac{N^3}{6\kappa}\,p^2 - p
    ,\qquad
    v(p) \sim
    1 + \frac{N^3}{3\kappa}\,p
    .
\end{equation}
Note that the asymptotic relations for $p(v)$ and $v(p)$ are equivalent, and that the Legendre transformation from Eq.~\eqref{eq:Legendre} holds true, as expected:

\begin{equation}\label{eq:Legendre-non-dim}
    \beta\Delta G \sim \beta\Delta A - pv
    \quad \text{for } N,M\gg 1
    .
\end{equation}
Also, note that these results also match that of the mechanically-treated system under small applied loads.
To arrive at similar asymptotic relations for $k'$, an additional approximation must be made.
Specifically, the incremental transition state stretch $\Delta\lambda^\ddagger\equiv\lambda^\ddagger-1$ is assumed to be small.
For the Morse potential, the nondimensional Morse parameter $\alpha=\ln(2)/\Delta\lambda^\ddagger$ is then assumed to be relatively large.
In the isometric ensemble, the asymptotic relation is

\begin{equation}\label{eq:k.isometric.asymptotic}
    k'(v) \sim
    \frac{\omega_0}{2\pi} \, e^{-\Delta\varepsilon^\ddagger + 3\kappa\Delta v\Delta\lambda^\ddagger/N^2}
    ,
\end{equation}
where $\omega_0\equiv\sqrt{u_0''/m}$ is the harmonic vibration frequency, i.e., the attempt frequency, and $\Delta\varepsilon^\ddagger$ is the nondimensional potential energy barrier to the transition state.
Detailed steps are shown in Appendix~\ref{sec:thermo-limit-asymp}.
For the specific case of the Morse potential, $u_0''=2a^2u_0$ and $\Delta\varepsilon^\ddagger=\varepsilon/4$.
In the isotensional ensemble, the asymptotic relation is

\begin{equation}\label{eq:k.isotensional.asymptotic}
    k'(p) \sim
    \frac{\omega_0}{2\pi} \, e^{-\Delta\varepsilon^\ddagger + Np\Delta\lambda^\ddagger}
    .
\end{equation}
Note that Eqs.~\eqref{eq:k.isometric.asymptotic} and \eqref{eq:k.isotensional.asymptotic} are equivalent, as shown by substituting in the asymptotic relations for $p(v)$ or $v(p)$ from Eqs.~\eqref{eq:asymptotic.AP} and \eqref{eq:asymptotic.GV}.
Since these relations only depend on the shape of the potential bottom and the location of the transition state (bond break), they are generalizable to many different potentials, including the ideal brittle potential \cite{marder1995fluctuations,marder1996statistical}.
Notably, these simplified asymptotic relations for $k'$ are analogous to Bell's model \cite{bell1978models}.
Eq.~\eqref{eq:k.isotensional.asymptotic} is of the form $k'(f)\propto e^{\beta f\Delta x^\ddagger}$, where $f\equiv NP$ is the effective force and $\Delta x^\ddagger\equiv q^\ddagger-b$ is the effective distance to the transition state, which is the form often attributed to Bell \cite{rief1998elastically,dudko2006intrinsic,silberstein2013modeling,silberstein2014modeling}.
Alternatively, when proporting the effective force $f$ to the material stress \cite{tehrani2017effect}, the form of Eq.~\eqref{eq:k.isotensional.asymptotic} matches the model of \citet{zhurkov1965kinetic}.
Finally, Eq.~\eqref{eq:k.isotensional.asymptotic} as a function of stress also bears a resemblance to the model of \citet{argon1973theory}, which was formulated to capture viscoplastic flow in glassy polymers \cite{boyce1988large,hasan1993investigation,wu1993improved,hasan1995constitutive,qi2005stress}.

The net rate of crack growth is defined as

\begin{equation}\label{eq:net-rate-isometric}
    k^\mathrm{net}(v) \equiv k'(v) - k''(v)
    ,
\end{equation}
where the rate of reforming the bond behind the crack tip $k''$ is given by Eq.~\eqref{eq:k.isometric}, after replacing $q_{N+1}$ with $q_N$ in the transition state configuration integral $Q_\mathrm{con}^\ddagger$ in Eq.~\eqref{eq:Q_con_ts}.
In the isometric ensemble (see Appendix~\ref{sec:thermo-limit-asymp.isometric}),

\begin{equation}\label{eq:k-net.isometric.asymptotic}
    k^\mathrm{net}(v) \sim
    \frac{\omega_0}{\pi} \, e^{-\Delta\varepsilon^\ddagger + 3\kappa\Delta v\Delta\lambda^\ddagger/N^2} \, \sinh\left[\frac{9\kappa(\Delta v)^2}{4N^4}\right]
    ,
\end{equation}
and in the isotensional ensemble (see Appendix~\ref{sec:thermo-limit-asymp.isotensional}),

\begin{equation}\label{eq:k-net.isotensional.asymptotic}
    k^\mathrm{net}(p) \sim
    \frac{\omega_0}{\pi} \, e^{-\Delta\varepsilon^\ddagger + Np\Delta\lambda^\ddagger} \, \sinh\left(\frac{N^2p^2}{4\kappa}\right)
    .
\end{equation}
Note that Eqs.~\eqref{eq:k-net.isometric.asymptotic} and \eqref{eq:k-net.isotensional.asymptotic} are again equivalent, as a result of invoking the thermodynamic limit, verified via Eqs.~\eqref{eq:asymptotic.AP} and \eqref{eq:asymptotic.GV}.
Also note that the net rates in Eqs.~\eqref{eq:k-net.isometric.asymptotic} and \eqref{eq:k-net.isotensional.asymptotic} contain the same Bell term obtained previously for the forward rates in Eqs.~\eqref{eq:k.isometric.asymptotic} and \eqref{eq:k.isotensional.asymptotic}.
In certain cases, such as a vanishingly small transition state displacement at a fixed nondimensional bending stiffness ($\Delta\lambda^\ddagger\to 0$), the net rate in Eq.~\eqref{eq:k-net.isotensional.asymptotic} is approximated as

\begin{equation}
    k^\mathrm{net}(p) \sim
    \frac{\omega_0}{\pi} \, e^{-\Delta\varepsilon^\ddagger} \, \sinh\left(\frac{N^2p^2}{4\kappa}\right)
    .
\end{equation}

Through calculating the energy release rate $R$ for the discrete system, it is possible to relate the net rate of crack growth $k^\mathrm{net}$ to continuum theories for crack growth.
In the thermodynamic limit, the resulting linear asymptotic relation for $p(v)$ in Eq.~\eqref{eq:asymptotic.GV} allows the compliance method to be utilized when obtaining the energy release rate \cite{zehnder2012fracture}.
If $\partial s = b^2 \partial N$ is the differential increase in area as the crack advances, the energy release rate $R$ is then

\begin{equation}
    R =
    \frac{P^2}{2}\frac{\partial}{\partial s}\left(\frac{\Delta V}{P}\right) =
    \frac{P^2}{2b^2}\frac{\partial}{\partial N}\left(\frac{\Delta V}{P}\right)
    .
\end{equation}
The nondimensional energy release rate $\beta Rb^2$ is then

\begin{equation}\label{eq:R-nondim}
    \beta R b^2 =
    \frac{p^2}{2} \frac{\partial}{\partial N}\left(\frac{\Delta v}{p}\right) =
    \frac{N^2p^2}{2\kappa}
    .
\end{equation}
It then becomes convenient to define the generalized force $f \equiv NP = \sqrt{2Rcb^2}$, which appears in Eq.~\eqref{eq:k-net.isotensional.asymptotic}, and then define $\Delta x^\ddagger\equiv b\Delta\lambda^\ddagger$.
Combining Eqs.~\eqref{eq:R-nondim} and $f=NP$, Eq.~\eqref{eq:k-net.isotensional.asymptotic} becomes

\begin{equation}\label{eq:k_net_macro}
    k^\mathrm{net} \sim
    \frac{\omega_0}{\pi}\, \exp\left(\frac{f\Delta x^\ddagger - \Delta u^\ddagger}{kT}\right) \sinh\left(\frac{Rb^2}{2kT}\right)
    .
\end{equation}
Note that multiplying the net rate of crack growth $k^\mathrm{net}$ by the atomic spacing $b$ gives the crack growth velocity, i.e., Eq.~\eqref{eq:knet}.
The only macroscopic parameters in Eq.~\eqref{eq:k_net_macro} are the energy release rate $R$ and the temperature $T$.
The atomic parameters in Eq.~\eqref{eq:k_net_macro} are the attempt frequency $\omega_0$, atomic spacing and bond length $b$, atomic bending stiffness $c$, potential energy barrier $\Delta u^\ddagger$, and transition state bond displacement $\Delta x^\ddagger$.
As such, Eq.~\eqref{eq:k_net_macro} constitutes a useful physically based relation for modeling subcritical crack growth experiments \cite{preston1935time,freiman2009environmentally,delrio2022eliciting}.
Eq.~\eqref{eq:k_net_macro} matches the form obtained by \citet{marder1996statistical} for small $\Delta v$.
The atomic parameters in Eq.~\eqref{eq:k_net_macro} could be determined with electronic structure calculations, but it may also be possible to calibrate a subset of the parameters using experimental results to examine microscopic properties.

A simpler relation for the net rate of crack growth in Eq.~\eqref{eq:k_net_macro} is obtained by neglecting the Bell term, yielding

\begin{equation}\label{eq:k_net_macro_simpler}
    k^\mathrm{net} \sim
    \frac{\omega_0}{\pi}\, \exp\left(-\frac{\Delta u^\ddagger}{kT}\right) \sinh\left(\frac{Rb^2}{2kT}\right)
    .
\end{equation}
Forms of this relation, as well as equivalent forms, have been obtained previously and successfully used in modeling subcritical crack growth experiments \cite{lawn1975atomistic,wiederhorn1980micromechanisms,cook1993kinetics,krausz1988fracture,michalske1983molecular,ciccotti2009stress,le2009subcritical,cook2019thermal,grutzik2022kinetic}.
One key difference between Eq.~\eqref{eq:k_net_macro_simpler} and past relations is $\omega_0/\pi$ instead of $2kT/h$, though the former is correct.
Observable results from classical formulations cannot depend on the Planck constant $h$, and its presence in the classical formulation of statistical thermodynamics is merely cosmetic and meant to nondimensionalize partition functions \cite{mcq,zwanzig2001nonequilibrium}.
Though a prefactor of $kT/h$ appears when computing transition state theory rates, $h$ always factors out \cite{zwanzig2001nonequilibrium}.
In any case, the more rigorous approach demonstrated here leading to Eq.~\eqref{eq:k_net_macro} or \eqref{eq:k_net_macro_simpler} validates the general form of similar relations obtained previously.
Further, the systematic set of simplifying assumptions made in the process confirms the validity of these relations in the subcritical regime.
Finally, note that Eq.~\eqref{eq:k_net_macro_simpler} bears a striking resemblance to models for viscoplastic flow in some polymers \cite{eyring1936viscosity,fotheringham1978role,richeton2005formulation,ames2009thermo,silberstein2010constitutive,silberstein2013modeling,narayan2021fracture} and bulk metallic glasses \cite{argon1979plastic,steif1982strain,schuh2007mechanical,henann2008constitutive,dubach2009constitutive}.
This resemblance suggests that subcritical crack growth and viscoplasticity could involve common mechanisms \cite{grutzik2022kinetic}.

The net rate of crack growth is recast in terms of nondimensional variables for proper parametric study.
Since the atomic bending stiffness included in the model represents effects from the bulk material, the nondimensional bending stiffness $\kappa$ may be considered a proxy for the nondimensional modulus $\beta Eb^3$.
After defining the nondimensional energy release rate $\Xi \equiv \beta Rb^2 = N^2p^2/2\kappa$ and the reference rate $k_\mathrm{ref}\equiv (\omega_0/\pi) e^{-\beta\Delta u^\ddagger}$, Eq.~\eqref{eq:k_net_macro} becomes

\begin{equation}
    k^\mathrm{net}(\Xi) \sim
    k_\mathrm{ref} \, e^{\Delta\lambda^\ddagger\sqrt{2\kappa\Xi}} \sinh\left(\frac{\Xi}{2}\right)
    ,
\end{equation}
and Eq.~\eqref{eq:k_net_macro_simpler} is similarly nondimensionalized as

\begin{equation}
    k^\mathrm{net}(\Xi) \sim
    k_\mathrm{ref} \, \sinh\left(\frac{\Xi}{2}\right)
    .
\end{equation}
Note that while $\Xi$ and $\kappa$ have both microscopic and macroscopic interpretations, $\Delta\lambda^\ddagger$ has only a microscopic interpretation (transition state bond stretch increment).

\clearpage

\section{Numerical results}\label{sec:results}

The asymptotic approach of evaluating the crack model system is now demonstrated.
In either thermodynamic ensemble, the rate of breaking the crack tip bond $k'$ is calculated as a function of load using the asymptotic approach developed in Sec.~\ref{sec:asymptotic}.
These calculations are repeated for increasing nondimensional bond energy $\varepsilon$ and compared with the results of Monte Carlo calculations (details in Appendix~\ref{sec:monte-carlo}).
The $L_2$ norm is utilized to compute the relative error $e$ between the asymptotic and Monte Carlo approaches while varying the nondimensional bond energy $\varepsilon$.
Since the scale of $k'$ increases many orders of magnitude while increasing $\varepsilon$, the logarithm of $k'$ is considered when computing the relative error $e$ rather than $k'$ directly.
Afterwards, the rate of breaking the crack tip bond $k'$ and the net rate of crack growth $k^\mathrm{net}$ are computed via the asymptotic approach while increasing the system size.
In both cases, the results are compared with the simplified analytic relations obtained in the thermodynamic limit in Sec.~\ref{sec:asymptotic.thermodynamic}.
The base parameters for the crack model system are $N=8$, $M=8$, $\alpha=1$ (which is $\Delta\lambda^\ddagger=\ln 2$), $\varepsilon=100$, and $\kappa=100$.
All calculations were completed using the \texttt{Python} package \texttt{statMechCrack} \cite{buchegrutzikstatmechcrack2022}, which acknowledges support from several other \texttt{Python} packages \cite{numpy,scipy,matplotlib}.

\begin{figure}[t]
    \begin{center}
        \includegraphics{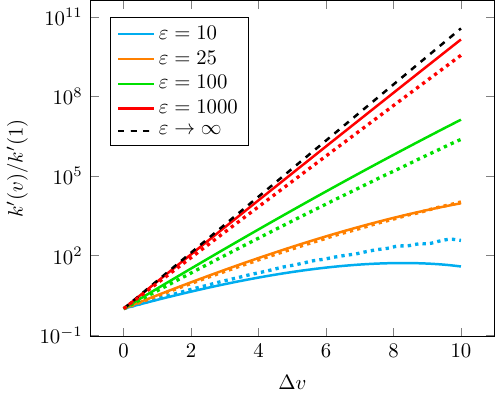}
    \end{center}
    \caption{\label{fig-isometric-rate}%
        The relative rate of breaking the crack tip bond as a function of the nondimensional end displacement, using the asymptotic approach (solid) and Monte Carlo calculations (dotted), for varying nondimensional bond energy $\varepsilon$.
    }
\end{figure}

\begin{figure}[t]
    \begin{center}
        \includegraphics{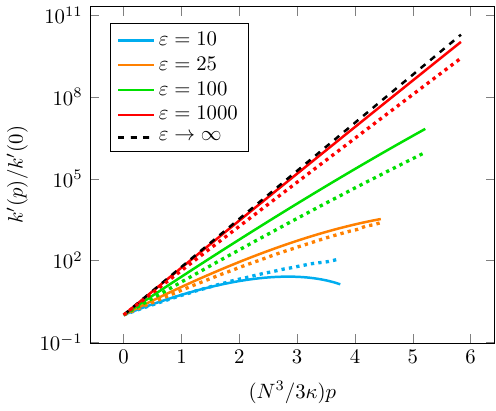}
    \end{center}
    \caption{\label{fig-isotensional-rate}%
        The relative rate of breaking the crack tip bond as a function of the rescaled nondimensional end force, using the asymptotic approach (solid) and Monte Carlo calculations (dotted), for varying nondimensional bond energy $\varepsilon$.
    }
\end{figure}

\subsection{Isometric ensemble}\label{sec:results.isometric}

In Fig.~\ref{fig-isometric-rate}, the rate of breaking the crack tip bond $k'(v)$ given by Eq.~\eqref{eq:k.isometric} is plotted relative to $k'(1)$ as a function of the nondimensional applied end displacement $\Delta v$.
The configuration integrals in Eq.~\eqref{eq:k.isometric} are asymptotically approximated in Sec.~\ref{sec:asymptotic.isometric} to analytically calculate $k'(v)$.
The Monte Carlo approach was also used to calculate $k'(v)$, as detailed in Appendix~\ref{sec:monte-carlo.isometric}, and is additionally plotted in Fig.~\ref{fig-isometric-rate}.
Both the asymptotic and Monte Carlo calculations were repeated while increasing the nondimensional bond energy $\varepsilon$, as shown in Fig.~\ref{fig-isometric-rate}.
For lower values of $\varepsilon$ (such as 10), the asymptotic approach tends to underestimate $k'$ significantly.
As $\varepsilon$ increases slightly (to 25), it appears to provide an excellent approximation, but this result is merely a coincidence of the curves passing over one another.
The is evident after $\varepsilon$ increases more (such as to 100), where the asymptotic approach then tends to overestimate $k'$.
When $\varepsilon$ becomes large (such as 1000), the asymptotic approach still overestimates $k'$, but the gap shrinks as $\varepsilon$ grows.
As $\varepsilon\to\infty$, both the asymptotic and Monte Carlo approaches of calculating $k'$ begin to match the asymptotic approach for $k'$ calculated using the reference system ($\boldsymbol{\lambda}=1$).

\begin{figure}[t]
    \begin{center}
        \includegraphics{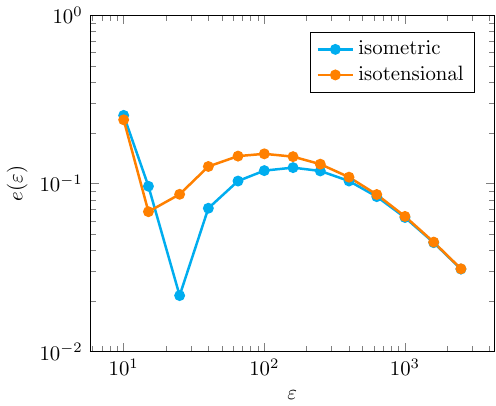}
    \end{center}
    \caption{\label{fig-rate-error}%
        The relative error $e$ when computing $k'$ using the asymptotic approach, as a function of the nondimensional bond energy $\varepsilon$, in either thermodynamic ensemble.
    }
\end{figure}

To make a more quantitative evaluation of the asymptotic approach of obtaining $k'(v)$, the relative error $e$ with respect to the Monte Carlo approach is calculated,

\begin{equation}\label{eq:rel-error.isometric}
    e(\varepsilon) =
    \sqrt{
        \frac{
            \int_1^{11} \ln\left[k(v) / k_\mathrm{m}(v)\right]^2 dv
        }{
            \int_1^{11} \ln\left[k_\mathrm{m}(v)\right]^2 \,dv
        }
    }
    ,
\end{equation}
where $k_\mathrm{m}$ is the result of the Monte Carlo calculations.
As shown in Fig.~\ref{fig-rate-error}, the relative error $e$ tends to decrease as the nondimensional bond energy $\varepsilon$ increases, apart from the region where the two approaches happen to overlap.
For large values of $\varepsilon$, the relative error steadily decreases at a log-log slope appearing to near $-1$, which supports the theory that the asymptotic approach becomes accurate for $\varepsilon\gg 1$.

\subsection{Isotensional ensemble}\label{sec:results.isotensional}

In Fig.~\ref{fig-isotensional-rate}, the rate of breaking the crack tip bond $k'(p)$ given by Eq.~\eqref{eq:k.isotensional} is plotted relative to $k'(0)$ as a function of the rescaled nondimensional applied force $N^3p/3\kappa$.
The configuration integrals in Eq.~\eqref{eq:k.isotensional} are asymptotically approximated in Sec.~\ref{sec:asymptotic.isotensional} to analytically calculate $k'(p)$.
The Monte Carlo approach was also used to calculate $k'(p)$, as detailed in Appendix~\ref{sec:monte-carlo.isotensional}, and is additionally plotted in Fig.~\ref{fig-isotensional-rate}.
Both the asymptotic and Monte Carlo calculations were repeated while increasing the nondimensional bond energy $\varepsilon$, as shown in Fig.~\ref{fig-isotensional-rate}.
For lower values of $\varepsilon$ (such as 10), the asymptotic approach tends to underestimate $k'$ significantly.
As $\varepsilon$ increases slightly (to 25), it appears to provide an excellent approximation, but this result is merely a coincidence of the curves passing over one another.
This is evident after $\varepsilon$ increases more (such as to 100), where the asymptotic approach then tends to overestimate $k'$.
When $\varepsilon$ becomes large (such as 1000), the asymptotic approach still overestimates $k'$, but the gap shrinks as $\varepsilon$ grows.
As $\varepsilon\to\infty$, both the asymptotic and Monte Carlo approaches of calculating $k'$ begin to match the asymptotic approach for $k'$ calculated using the reference system ($\boldsymbol{\lambda}=1$).

\begin{equation}\label{eq:rel-error.isotensional}
    e(\varepsilon) =
    \sqrt{
        \frac{
            \int_0^{10} \ln\left[k(p) / k_\mathrm{m}(p)\right]^2 dp
        }{
            \int_0^{10} \ln\left[k_\mathrm{m}(p)\right]^2 \,dp
        }
    }
    ,
\end{equation}
where $k_\mathrm{m}$ is the result of the Monte Carlo calculations.
As shown in Fig.~\ref{fig-rate-error}, the relative error $e$ tends to decrease as the nondimensional bond energy $\varepsilon$ increases, apart from the region where the two approaches happen to overlap.
For large values of $\varepsilon$, the relative error steadily decreases at a log-log slope appearing to near $-1$, which supports the theory that the asymptotic approach becomes accurate for $\varepsilon\gg 1$.

\subsection{Thermodynamic limit}\label{sec:results.thermodynamic}

\begin{figure}[t]
    \begin{center}
        \includegraphics{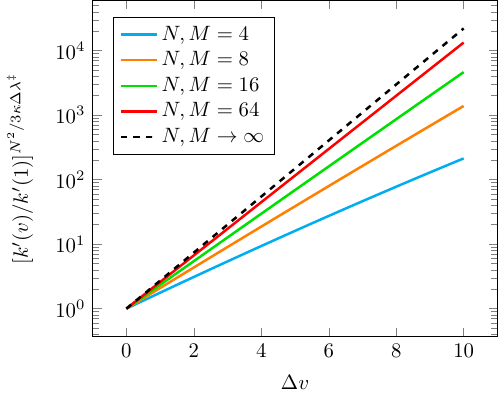}
    \end{center}
    \caption{\label{fig-limit-rate}%
        The rescaled relative rate of breaking the crack tip bond as a function of the nondimensional end displacement, using the asymptotic approach, for increasing system size.
    }
\end{figure}

The rescaled rate of breaking the crack tip bond is plotted in Fig.~\ref{fig-limit-rate} as a function of the nondimensional applied end displacement $\Delta v$, calculated using the asymptotic approach and repeated as the system size ($N,M$) increases.
The rescaling of $k'(v)$ in Fig.~\ref{fig-limit-rate} corresponds to the simplified relation for $k'(v)$ obtained in the thermodynamic limit, given by Eq.~\eqref{eq:k.isometric.asymptotic} or equivalently given by Eq.~\eqref{eq:k.isotensional.asymptotic}.
Since obtaining these relations for $k'(v)$ involved making approximations based on small $\Delta\lambda^\ddagger$, from here on the model uses $\Delta\lambda^\ddagger=0.1$ (which is $\alpha=10\ln 2$) in addition to previously incorporated assumptions ($\varepsilon\gg 1$ and $N,M\gg 1$).
As shown in Fig.~\ref{fig-limit-rate}, as the system becomes large the rate of breaking the crack tip bond $k'$ approaches the simpler relation in Eq.~\eqref{eq:k.isometric.asymptotic}.
To reiterate, Eq.~\eqref{eq:k.isometric.asymptotic} would not necessarily succeed if the crack tip bond potential was not steep or if the displacement required to break the bond was not small.
As the applied load becomes large, $k'$ predictably diverges from Eq.~\eqref{eq:k.isometric.asymptotic}, even for large system sizes \cite{buche2020statistical}.
Note that the same results in Fig.~\ref{fig-limit-rate} were calculated in the isotensional ensemble, but these curves exactly matched those from the isometric ensemble.
Though not apparent at the outset, this match could mean that certain results of the reference system -- when governed by purely harmonic potentials -- can be treated independent of the thermodynamic ensemble.
This effect is also somewhat evident in Fig.~\ref{fig-rate-error}, where the performance of the asymptotic approach becomes independent of ensemble as $\varepsilon$ increases.

\begin{figure}[t]
    \begin{center}
        \includegraphics{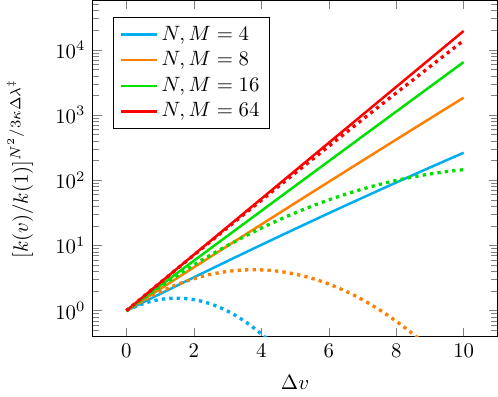}
    \end{center}
    \caption{\label{fig-forward-rev-net-rate}%
        The rescaled relative rates of breaking the crack tip (solid) and reforming behind the crack tip (dotted) as a function of the nondimensional end displacement, calculated using the asymptotic approach, for increasing system size.
    }
\end{figure}

The rescaled rate of breaking the crack tip bond $k'$ is plotted again in Fig.~\ref{fig-forward-rev-net-rate} along with the rate of reforming the bond behind the crack tip $k''$.
As the nondimensional applied end displacement  $\Delta v$ increases, the forward rate $k'$ eventually dominates the reverse rate $k''$, producing a net rate of crack growth $k^\mathrm{net}\equiv k'-k''$.
As the system size ($N,M$) increases, this domination appears to diminsh, but this is simply a result of the rescaling necessary to fit each curve in Fig.~\ref{fig-forward-rev-net-rate}.
As will be demonstrated shortly, the net rate of crack growth does indeed converge to a nontrivial curve as the system size becomes large.

\begin{figure}[t]
    \begin{center}
        \includegraphics{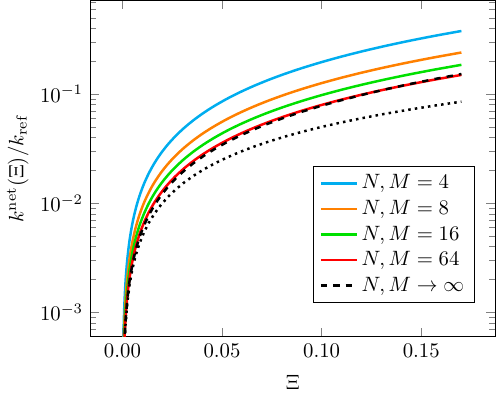}
    \end{center}
    \caption{\label{fig-limit-net-rate}%
        The relative net rate of crack growth as a function of the nondimensional energy release rate, calculated using the asymptotic approach, for increasing system size.
    }
\end{figure}

The relative net rate of crack growth $k^\mathrm{net}$ is plotted in Fig.~\ref{fig-limit-net-rate} relative to the reference rate $k_\mathrm{ref}\equiv (\omega_0/\pi) e^{-\beta\Delta u^\ddagger}$ as a function of the nondimensional energy release rate $\Xi=N^2p^2/2\kappa$, calculated using the asymptotic approach and repeated as the system size ($N,M$) increases.
Note that the black dashed line represents Eq.~\eqref{eq:k_net_macro}, and the black dotted line represents Eq.~\eqref{eq:k_net_macro_simpler}.
Fig.~\ref{fig-limit-net-rate} shows that the net rate of crack growth $k^\mathrm{net}$ approaches the simplified relation given by Eq.~\eqref{eq:k_net_macro} as the system becomes large.
In a similar way as before, this asymptotic behavior will eventually no longer hold when the applied load becomes large enough.
For intermediate to large nondimensional energy release rates $\Xi$, the asymptotic approach for $k^\mathrm{net}$ will diverge from Eq.~\eqref{eq:k_net_macro}.
This divergence means that Eq.~\eqref{eq:k_net_macro} is typically valid only for relatively small energy release rates, i.e., subcritical crack growth, which is primarily due to the small bond stretch assumption necessary to arive at Eq.~\eqref{eq:k_net_macro}.
Fig.~\ref{fig-limit-net-rate} further shows that for nonzero albeit small $\Delta\lambda^\ddagger$, the Bell-like term differentiating Eq.~\eqref{eq:k_net_macro} from Eq.~\eqref{eq:k_net_macro_simpler} contributes significantly to $k^\mathrm{net}$ and allows the correct thermodynamic limit relation for $k^\mathrm{net}$ to be obtained.
It is then more accurate to use the relation for the net rate of crack growth $k^\mathrm{net}$ in Eq.~\eqref{eq:k_net_macro} than the the simpler relation in Eq.~\eqref{eq:k_net_macro_simpler} used previously \cite{lawn1975atomistic,wiederhorn1980micromechanisms,cook1993kinetics,krausz1988fracture,michalske1983molecular,ciccotti2009stress,le2009subcritical,cook2019thermal,grutzik2022kinetic}.

\subsection{Subcritical crack growth experiments}\label{sec:results.scg}

\begin{figure}[t]
    \begin{center}
        \includegraphics{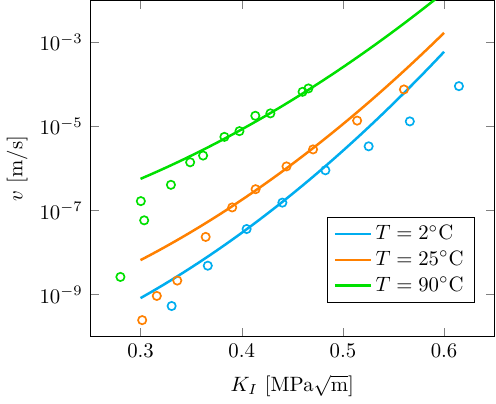}
    \end{center}
    \caption{\label{fig-scg}%
        The crack growth velocity as a function of the stress intensity factor, calculated using the simplified asymptotic relation (solid), and the experimental measurements (circles) from SLS glass in water \cite{wiederhorn1970stress}, for increasing temperature.
    }
\end{figure}

Eq.~\eqref{eq:k_net_macro} is now used to model subcritical crack growth experiments from \citet{wiederhorn1970stress} involving soda-lime silicate glass in water at varying temperature.
A mode-I stress intensity factor $K_I=\sqrt{RE}$ is applied, where the modulus is $E=73$~GPa \cite{grutzik2022kinetic}.
The crack growth velocity is given by the rate $k^\mathrm{net}$ multiplied by the length $b$, i.e., Eq.~\eqref{eq:knet}.
The attempt frequency $\omega_0=3.3\times 10^{13}\,\mathrm{s}^{-1}$ is obtained from the median wave number (1100~cm$^{-1}$) of the Si-O-Si stretching mode in silicate glasses via infrared spectroscopy \cite{borrajo2004influence}.
The bond length $b=1.6$~\AA\ and crack tip bond transition state energy $\Delta u^\ddagger=1.22\times 10^{-12}$~J are from existing calibrations for soda-lime silicate glass in water \cite{grutzik2022kinetic}.
The transition state bond displacement is then calibrated to be $\Delta x^\ddagger=0.04b$, which is realistic compared to reactive molecular dynamics calculations that estimate $\Delta x$ \cite{yue2015tribochemical,yeon2016reaxff}.
The calibration of the transition state bond displacement $\Delta x^\ddagger$ to the experimental data demonstrates how macroscopic experiments may be used with this approach to examine microscopic properties.
The results are shown in Fig.~\ref{fig-scg}, where the subcritical crack growth velocities given by Eq.~\eqref{eq:knet} provide reasonable predictions of the experimentally measured velocities over the range of applied stress intensity factors and several temperatures.
Note that Eq.~\eqref{eq:knet} underestimates the stress intensity needed to reach higher velocities, due to it not including the inhibiting effects of finite water diffusion rates \cite{grutzik2022kinetic}.
Also note that Eq.~\eqref{eq:knet} overpredicts the velocity for small stress intensity since it does not account for the threshold effects within this material \cite{fett2005interpretation,grutzik2022kinetic}.

\clearpage

\section{Conclusion}\label{sec:conclusion}

The principles of statistical thermodynamics have been applied to an idealized particle-based model of a crack under an applied load.
In both the isometric and isotensional ensembles, the partition function was formulated for the model system to derive thermodynamic quantities, such as the free energy, and the kinetic rate of breaking the crack tip bond.
An asymptotic approach was utilized to obtain analytic relations for the rate of breaking the crack tip bond, valid in the limit that the bond potentials ahead of and including the crack tip are steep.
These asymptotic relations were developed in either thermodynamic ensemble, before being verified numerically with respect to Monte Carlo calculations.
Simplified analytic relations were obtained and verified for the rate of breaking the crack tip bond and the net rate of crack growth, valid in the thermodynamic limit of large system size and the subcritical regime.
The analytic relation for the net rate of crack growth ultimately obtained here offers an effective, practical, and physical method for modeling subcritical crack growth experiments, which was explicitly demonstrated by modeling subcritical crack growth in soda-lime silicate glass.
In future work, it would be useful to calculate the rate of breaking consecutive bonds ahead of the crack tip and model the subcritical to critical crack growth transition.
To account for different regimes of subcritical crack growth, future work should also consider generalizing the model system to include chemical interactions at the crack tip.
Finally, it could be important in future work to include inhomgeneity in the system to model start-stop behavior and related phenomena.

\begin{acknowledgments}
This work was supported by the Laboratory Directed Research and Development program at Sandia National Laboratories under Project No. 222398.
Sandia National Laboratories is a multi-mission laboratory managed and operated by National Technology and Engineering Solutions of Sandia, LLC., a wholly owned subsidiary of Honeywell International, Inc., for the U.S. Department of Energy's National Nuclear Security Administration under Contract No. DE-NA0003525.
Any subjective views or opinions expressed in the paper do not necessarily represent the views of the U.S. Department of Energy or the U.S. Government.
The U.S. Government retains and the publisher, by accepting the article for publication, acknowledges that the U.S. Government retains a nonexclusive, paid-up, irrevocable, world-wide license to publish or reproduce the published form of this manuscript, or allow others to do so, for U.S. Government purposes.
\end{acknowledgments}

\appendix
\section{Reference system calculations}\label{sec:ref-sys}

In Secs.~\ref{sec:asymptotic.isometric} and \ref{sec:asymptotic.isotensional}, the statistical thermodynamics of the full model system are asymptotically approximated in terms of the reference system.
The reference system is equivalent to the statistical mechanical treatment of the discrete representation of a linear elastic slender beam with a fixed end displacement (isometric) or end force (isotensional).
The configuration integral for this reference system can be computed analytically in either thermodynamic ensemble, as shown here.

\subsection{Isometric ensemble}\label{sec:ref-sys.isometric}

Here the integral in Eq.~\eqref{eq:Q_0_con}, defined as

\begin{equation}
    I(v,\boldsymbol{\lambda}) \equiv
    \int ds_1\cdots ds_N \ e^{-\beta U_{00}(\mathbf{s})}
    ,
\end{equation}
will be computed analytically.
$U_{00}$ is given by Eq.~\eqref{eq:betaU_00}, and can be rewritten in the quadratic form

\begin{equation}
    \beta U_{00}(\mathbf{s}) =
    \frac{1}{2}\,\mathbf{s}^T\cdot\mathbf{H}\cdot\mathbf{s} - \mathbf{g}^T\cdot\mathbf{s} + f
    ,
\end{equation}
where the Hessian $\mathbf{H}$ is given by Eq.~\eqref{eq:H.isometric}, and $\mathbf{g}$ and $f$ are given by Eq.~\eqref{eq:gf.isometric}.
The integral is now

\begin{equation}
    I(v,\boldsymbol{\lambda}) =
    e^{-f}\int ds_1\cdots ds_N \ e^{-\tfrac{1}{2}\mathbf{s}^T\cdot\mathbf{H}\cdot\mathbf{s} + \mathbf{g}^T\cdot\mathbf{s}}
    ,
\end{equation}
which takes the same form as the path integral in free-scalar relativistic quantum field theory \cite{zee2010quantum}, from which we have the exact result

\begin{equation}
    I(v,\boldsymbol{\lambda}) =
    \sqrt{\frac{(2\pi)^N}{\det\mathbf{H}}} \ e^{\tfrac{1}{2}\mathbf{g}^T\cdot\mathbf{H}^{-1}\cdot\mathbf{g} - f}
    ,
\end{equation}
which leads directly to $Q_\mathrm{0,con}(v,\boldsymbol{\lambda})$ in Eq.~\eqref{eq:Q_0_con_result}.

\subsection{Isotensional ensemble}\label{sec:ref-sys.isotensional}

Here the integral in Eq.~\eqref{eq:Z_0_con}, defined as

\begin{equation}
    I(p,\boldsymbol{\lambda}) \equiv
    \int ds_0\cdots ds_N \ e^{-\beta\Pi_{00}(\mathbf{s})}
    ,
\end{equation}
will be computed analytically.
$\beta\Pi_{00}\equiv \beta U_{00} - ps_0$ can be rewritten in the quadratic form

\begin{equation}
    \beta\Pi_{00}(\mathbf{s}) =
    \frac{1}{2}\,\mathbf{s}^T\cdot\mathbf{H}\cdot\mathbf{s} - \mathbf{g}^T\cdot\mathbf{s} + f
    ,
\end{equation}
where the Hessian $\mathbf{H}$ is given by Eq.~\eqref{eq:H.isotensional}, and $\mathbf{g}$ and $f$ are given by Eq.~\eqref{eq:gf.isotensional}.
The integral is now

\begin{equation}
    I(p,\boldsymbol{\lambda}) =
    e^{-f}\int ds_0\cdots ds_N \ e^{-\tfrac{1}{2}\mathbf{s}^T\cdot\mathbf{H}\cdot\mathbf{s} + \mathbf{g}^T\cdot\mathbf{s}}
    ,
\end{equation}
which has the exact result

\begin{equation}
    I(p,\boldsymbol{\lambda}) =
    \sqrt{\frac{(2\pi)^{N+1}}{\det\mathbf{H}}} \ e^{\tfrac{1}{2}\mathbf{g}^T\cdot\mathbf{H}^{-1}\cdot\mathbf{g} - f}
    ,
\end{equation}
which leads directly to $Z_\mathrm{0,con}(p,\boldsymbol{\lambda})$ in Eq.~\eqref{eq:Z_0_con_result}.

\section{Thermodynamic limit calculations}\label{sec:thermo-limit-asymp}

The analytic and asymptotically correct (for $\varepsilon\gg 1$) relations obtained in Sec.~\ref{sec:asymptotic.isometric} and Sec.~\ref{sec:asymptotic.isotensional} are reconsidered in the thermodynamic limit of large system size $(N,M\gg 1)$ to obtain the relations in Sec.~\ref{sec:asymptotic.thermodynamic}.
For the rate of breaking the crack tip bond and the net rate of crack growth, additional approximations are made when the incremental transition state stretch is small ($\Delta\lambda^\ddagger \ll 1$).
Though the results in either ensemble are equivalent due to the thermodynamic limit, the analysis is repeated in both ensembles for completeness.

\subsection{Isometric ensemble}\label{sec:thermo-limit-asymp.isometric}

As the system becomes large and the applied nondimensional displacement $\Delta v$ remains comparably small, bond stretching ahead of the crack tip becomes negligible ($\hat{\lambda}_j\sim 1$).
In this case ($\varepsilon\gg 1$ and $N,M\gg 1$), the asymptotic relation for $Q_\mathrm{con}$ in Eq.~\eqref{eq:Q_con_asymp} becomes

\begin{equation}\label{eq:Q_con_asymp_large}
    Q_\mathrm{con}(v) \sim
    Q_\mathrm{0,con}(v,1,1) \prod_{j=1}^M
        \sqrt{\frac{2\pi}{\beta u''(1)}}
    .
\end{equation}
When computing $\beta\Delta A(v)$ using Eq.~\eqref{eq:betaDeltaAandp}, the product term above cancels, leaving only the reference system configuration integral $Q_\mathrm{0,con}$ for $\lambda_1$$=\lambda_2=1$.
Using Eqs.~\eqref{eq:Q_0_con_result}--\eqref{eq:gf.isometric}, in this case the reference system configuration integral is

\begin{equation}\label{eq:Q_con_ref_simp}
    Q_\mathrm{0,con}(v, 1, 1) =
    \sqrt{\frac{(2\pi)^N}{\det\mathbf{H}}} \ e^{-3\kappa(\Delta v)^2/2N^3}
    .
\end{equation}
Using Eq.~\eqref{eq:betaDeltaAandp}, the nondimensional relative Helmholtz free energy and the nondimensional force have the asymptotic relations, valid for $\varepsilon\gg 1$ and $N,M\gg 1$,

\begin{equation}
    \beta\Delta A(v) \sim
    \frac{3\kappa}{2N^3}\left(\Delta v\right)^2
    ,\qquad
    p(v) \sim
    \frac{3\kappa}{N^3}\,\Delta v
    ,
\end{equation}
which is Eq.~\eqref{eq:asymptotic.AP}.
The rate of breaking the crack tip bond $k'$ is given by Eq.~\eqref{eq:k.isometric}, where the transition state configuration integral $Q_\mathrm{con}^\dagger$ is asymptotically given by Eq.~\eqref{eq:Q_con_asymp} with $\lambda_1=\lambda^\ddagger$.
For $\Delta\lambda^\ddagger \ll 1$, in addition to the previously invoked conditions ($\varepsilon\gg 1$ and $N,M\gg 1$), only the crack tip bond stretch is nonnegligible and $Q_\mathrm{con}^\dagger$ can be asymptotically approximated similar to $Q_\mathrm{con}$ in Eq.~\eqref{eq:Q_con_asymp_large},

\begin{equation}\label{eq:Q_con_TS_asymp_large}
    Q_\mathrm{con}^\dagger(v) \sim
    Q_\mathrm{0,con}(v,\lambda^\ddagger,1) \, e^{-\Delta\varepsilon^\ddagger} \prod_{j=2}^M \sqrt{\frac{2\pi}{\beta u''(1)}}
    ,
\end{equation}
where $\Delta\varepsilon^\ddagger\equiv\beta u(\lambda^\ddagger)-\beta u(1)$.
Eq.~\eqref{eq:k.isometric} then yields

\begin{equation}\label{eq:k_asymp_1}
    k'(v) \sim
    \frac{\omega_0}{2\pi} \, e^{-\Delta\varepsilon^\ddagger} \, \frac{Q_\mathrm{0,con}(v,\lambda^\ddagger,1)}{Q_\mathrm{0,con}(v,1,1)}
    ,
\end{equation}
where $\omega_0\equiv\sqrt{u''(1)/m}$.
When $\Delta\lambda^\ddagger$ is small, squares and higher powers of $\Delta\lambda^\ddagger$ are negligible, so the reference system transition state configuration integral, using Eqs.~\eqref{eq:Q_0_con_result}--\eqref{eq:gf.isometric} with $\lambda_1=\lambda^\ddagger$ and $\lambda_2=1$, becomes

\begin{equation}\label{eq:Q_con_ref_TS_simp}
    Q_\mathrm{0,con}(v, \lambda^\ddagger, 1) \sim
    \sqrt{\frac{(2\pi)^N}{\det\mathbf{H}}} \ e^{-3\kappa\Delta v(\Delta v-2\Delta\lambda^\ddagger)/2N^3}
    .
\end{equation}
Combining Eqs.~\eqref{eq:Q_con_ref_simp}, \eqref{eq:k_asymp_1}, and \eqref{eq:Q_con_ref_TS_simp}, the rate of breaking the crack tip bond has the asymptotic relation, valid for $\varepsilon\gg 1$, $N,M\gg 1$, and $\Delta\lambda^\ddagger \ll 1$,

\begin{equation}\label{eq:k.isometric.asymptotic.appendix}
    k'(v) \sim
    \frac{\omega_0}{2\pi} \, e^{-\Delta\varepsilon^\ddagger + 3\kappa\Delta v\Delta\lambda^\ddagger/N^2}
    ,
\end{equation}
which is Eq.~\eqref{eq:k.isometric.asymptotic}.
To compute the net rate of crack growth in Eq.~\eqref{eq:net-rate-isometric}, first the reverse rate must be written using Eq.~\eqref{eq:k.isometric}, which is

\begin{equation}\label{eq:k-rev.isometric}
    k''(N,M) =
    \sqrt{\frac{1}{2\pi m \beta}} \ \frac{Q_\mathrm{con}^\ddagger(N-1,M+1)}{Q_\mathrm{con}(N,M)}
    .
\end{equation}
As the system becomes large, the rate of breaking the crack tip bond is approximately the same as subsequently breaking the next bond.
The same holds for the rate of reforming the bond behind the crack tip, so $k''(N,M)\sim k''(N+1,M-1)$ for $N,M\gg 1$, which then means

\begin{equation}\label{eq:Q_con_ratio_approx}
    \frac{Q_\mathrm{con}^\ddagger(N-1,M+1)}{Q_\mathrm{con}(N,M)} \sim
    \frac{Q_\mathrm{con}^\ddagger(N,M)}{Q_\mathrm{con}(N+1,M-1)}
    .
\end{equation}
Additionally, $M-1\sim M$ for $M\gg 1$, so Eq.~\eqref{eq:net-rate-isometric} can now be approximated using Eqs.~\eqref{eq:k.isometric}, \eqref{eq:k-rev.isometric}, and \eqref{eq:Q_con_ratio_approx} as

\begin{equation}\label{eq:k-net-1.isometric}
    k^\mathrm{net} \sim
    \frac{Q_\mathrm{con}^{\ddagger}(N)}{\sqrt{2\pi m\beta}} \left(\frac{1}{Q_\mathrm{con}(N)} - \frac{1}{Q_\mathrm{con}(N\mathrm{+}1)}\right)
    .
\end{equation}
The configuration integral is related to the Helmholtz free energy $A$ via Eq.~\eqref{eq:AP}.
Using the asymptotic relations for $\beta\Delta A$ in Eq.~\eqref{eq:asymptotic.AP} under $N\gg 1$,

\begin{equation}
    \begin{aligned}
        \beta A(N) - \beta A(N+1) \sim& \
            \frac{\kappa}{2}\left(\frac{3\Delta v}{N^2}\right)^2
        ,\\
        \beta A(N) + \beta A(N+1) \sim& \
            2\beta A(N)
        ,
    \end{aligned}
\end{equation}
which then allows Eq.~\eqref{eq:k-net-1.isometric} to be rewritten as

\begin{equation}
    k^\mathrm{net} \sim
    \sqrt{\frac{2}{\pi m \beta}} \ \frac{Q_\mathrm{con}^{\ddagger}(N)}{Q_\mathrm{con}(N)}  \,\sinh\left[\frac{9\kappa(\Delta v)^2}{4N^4}\right]
    .
\end{equation}
Looking back to Eq.~\eqref{eq:k.isometric}, the terms outside the hyperbolic sine amount to $2k'$, where $k'$ is given by Eq.~\eqref{eq:k.isometric.asymptotic.appendix}.
Therefore the net rate of crack growth has the asymptotic relation, valid for $\varepsilon\gg 1$, $N,M\gg 1$, and $\Delta\lambda^\ddagger \ll 1$,

\begin{equation}
    k^\mathrm{net} \sim
    \frac{\omega_0}{\pi} \, e^{-\Delta\varepsilon^\ddagger + 3\kappa\Delta v\Delta\lambda^\ddagger/N^2} \, \sinh\left[\frac{9\kappa(\Delta v)^2}{4N^4}\right]
    ,
\end{equation}
which is Eq.~\eqref{eq:k-net.isometric.asymptotic}.

\subsection{Isotensional ensemble}\label{sec:thermo-limit-asymp.isotensional}

As the system becomes large and the applied nondimensional force $p$ remains comparably small, bond stretching ahead of the crack tip becomes negligible ($\hat{\lambda}_j\sim 1$).
In this case ($\varepsilon\gg 1$ and $N,M\gg 1$), the asymptotic relation for $Z_\mathrm{con}$ in Eq.~\eqref{eq:Z_con_asymp} becomes

\begin{equation}\label{eq:Z_con_asymp_large}
    Z_\mathrm{con}(p) \sim
    Z_\mathrm{0,con}(p,1,1) \prod_{j=1}^M
        \sqrt{\frac{2\pi}{\beta u''(1)}}
    .
\end{equation}
When computing $\beta\Delta G(p)$ using Eq.~\eqref{eq:betaDeltaGandv}, the product term above cancels, leaving only the reference system configuration integral $Z_\mathrm{0,con}$ for $\lambda_1$$=\lambda_2=1$.
Using Eqs.~\eqref{eq:Z_0_con_result}--\eqref{eq:gf.isotensional}, in this case the reference system configuration integral is

\begin{equation}\label{eq:Z_con_ref_simp}
    Z_\mathrm{0,con}(p, 1, 1) =
    \sqrt{\frac{(2\pi)^N}{\det\mathbf{H}}} \ e^{N^3p^2/6\kappa+p}
    .
\end{equation}
Using Eq.~\eqref{eq:betaDeltaGandv}, the nondimensional relative Helmholtz free energy and the nondimensional force have the asymptotic relations, valid for $\varepsilon\gg 1$ and $N,M\gg 1$,

\begin{equation}
    \beta\Delta G(p) \sim
    -\frac{N^3}{6\kappa}\,p^2 - p
    ,\qquad
    v(p) \sim
    1 + \frac{N^3}{3\kappa}\,p
    ,
\end{equation}
which is Eq.~\eqref{eq:asymptotic.GV}.
The rate of breaking the crack tip bond $k'$ is given by Eq.~\eqref{eq:k.isotensional}, where the transition state configuration integral $Z_\mathrm{con}^\dagger$ is asymptotically given by Eq.~\eqref{eq:Z_con_asymp} with $\lambda_1=\lambda^\ddagger$.
For $\Delta\lambda^\ddagger \ll 1$, in addition to the previously invoked conditions ($\varepsilon\gg 1$ and $N,M\gg 1$), only the crack tip bond stretch is nonnegligible and $Z_\mathrm{con}^\dagger$ can be asymptotically approximated similar to $Z_\mathrm{con}$ in Eq.~\eqref{eq:Z_con_asymp_large},

\begin{equation}\label{eq:Z_con_TS_asymp_large}
    Z_\mathrm{con}^\dagger(p) \sim
    Z_\mathrm{0,con}(p,\lambda^\ddagger,1) \, e^{-\Delta\varepsilon^\ddagger} \prod_{j=2}^M \sqrt{\frac{2\pi}{\beta u''(1)}}
    ,
\end{equation}
where $\Delta\varepsilon^\ddagger\equiv\beta u(\lambda^\ddagger)-\beta u(1)$.
Eq.~\eqref{eq:k.isotensional} then yields

\begin{equation}\label{eq:k_asymp_2}
    k'(p) \sim
    \frac{\omega_0}{2\pi} \, e^{-\Delta\varepsilon^\ddagger} \, \frac{Z_\mathrm{0,con}(p,\lambda^\ddagger,1)}{Z_\mathrm{0,con}(p,1,1)}
    ,
\end{equation}
where $\omega_0\equiv\sqrt{u''(1)/m}$.
When $\Delta\lambda^\ddagger$ is small, squares and higher powers of $\Delta\lambda^\ddagger$ are negligible, so the reference system transition state configuration integral, using Eqs.~\eqref{eq:Z_0_con_result}--\eqref{eq:gf.isotensional} with $\lambda_1=\lambda^\ddagger$ and $\lambda_2=1$, becomes

\begin{equation}\label{eq:Z_con_ref_TS_simp}
    Z_\mathrm{0,con}(p, \lambda^\ddagger, 1) \sim
    \sqrt{\frac{(2\pi)^{N+1}}{\det\mathbf{H}}} \ e^{N^3p^2/6\kappa+p+Np\Delta\lambda^\ddagger}
    .
\end{equation}
Combining Eqs.~\eqref{eq:Z_con_ref_simp}, \eqref{eq:k_asymp_2}, and \eqref{eq:Z_con_ref_TS_simp}, the rate of breaking the crack tip bond has the asymptotic relation, valid for $\varepsilon\gg 1$, $N,M\gg 1$, and $\Delta\lambda^\ddagger \ll 1$,

\begin{equation}\label{eq:k.isotensional.asymptotic.appendix}
    k'(p) \sim
    \frac{\omega_0}{2\pi} \, e^{-\Delta\varepsilon^\ddagger + Np\Delta\lambda^\ddagger}
    ,
\end{equation}
which is Eq.~\eqref{eq:k.isotensional.asymptotic}.
To compute the net rate of crack growth, first the reverse rate must be written using Eq.~\eqref{eq:k.isotensional}, which is

\begin{equation}\label{eq:k-rev.isotensional}
    k''(N,M) =
    \sqrt{\frac{1}{2\pi m \beta}} \ \frac{Z_\mathrm{con}^\ddagger(N-1,M+1)}{Z_\mathrm{con}(N,M)}
    .
\end{equation}
As the system becomes large, the rate of breaking the crack tip bond is approximately the same as subsequently breaking the next bond.
The same holds for the rate of reforming the bond behind the crack tip, so $k''(N,M)\sim k''(N+1,M-1)$ for $N,M\gg 1$, which then means

\begin{equation}\label{eq:Z_con_ratio_approx}
    \frac{Z_\mathrm{con}^\ddagger(N-1,M+1)}{Z_\mathrm{con}(N,M)} \sim
    \frac{Z_\mathrm{con}^\ddagger(N,M)}{Z_\mathrm{con}(N+1,M-1)}
    .
\end{equation}
Additionally, $M-1\sim M$ for $M\gg 1$, so $k^\mathrm{net}$ can now be approximated using Eqs.~\eqref{eq:k.isotensional}, \eqref{eq:k-rev.isotensional}, and \eqref{eq:Z_con_ratio_approx} as

\begin{equation}\label{eq:k-net-2.isotensional}
    k^\mathrm{net} \sim
    \frac{Z_\mathrm{con}^{\ddagger}(N)}{\sqrt{2\pi m\beta}} \left(\frac{1}{Z_\mathrm{con}(N)} - \frac{1}{Z_\mathrm{con}(N\mathrm{+}1)}\right)
    .
\end{equation}
The configuration integral is related to the Gibbs free energy $G$ via Eq.~\eqref{eq:GV}.
Using the asymptotic relations for $\beta\Delta G$ in Eq.~\eqref{eq:asymptotic.GV} under $N\gg 1$,

\begin{equation}
    \begin{aligned}
        \beta G(N) - \beta G(N+1) \sim& \
            \frac{N^2p^2}{2\kappa}
        ,\\
        \beta G(N) + \beta G(N+1) \sim& \
            2\beta G(N)
        ,
    \end{aligned}
\end{equation}
which then allows Eq.~\eqref{eq:k-net-2.isotensional} to be rewritten as

\begin{equation}
    k^\mathrm{net} \sim
    \sqrt{\frac{2}{\pi m \beta}} \ \frac{Z_\mathrm{con}^{\ddagger}(N)}{Z_\mathrm{con}(N)}  \, \sinh\left(\frac{N^2p^2}{4\kappa}\right)
    .
\end{equation}
Looking back to Eq.~\eqref{eq:k.isotensional}, the terms outside the hyperbolic sine amount to $2k'$, where $k'$ is given by Eq.~\eqref{eq:k.isotensional.asymptotic.appendix}.
Therefore the net rate of crack growth has the asymptotic relation, valid for $\varepsilon\gg 1$, $N,M\gg 1$, and $\Delta\lambda^\ddagger \ll 1$,

\begin{equation}
    k^\mathrm{net} \sim
    \frac{\omega_0}{\pi} \, e^{-\Delta\varepsilon^\ddagger + Np\Delta\lambda^\ddagger} \, \sinh\left(\frac{N^2p^2}{4\kappa}\right)
    ,
\end{equation}
which is Eq.~\eqref{eq:k-net.isotensional.asymptotic}.

\section{Monte Carlo calculations}\label{sec:monte-carlo}

Metropolis-Hastings Markov chain Monte Carlo calculations \cite{haile1992molecular} were performed \cite{buchegrutzikstatmechcrack2022} to verify the results (shown in Sec.~\ref{sec:results}) of the asymptotic approach.
Any ensemble average involved with this Monte Carlo approach can only calculate free energies and transition state theory reaction rates relative to a reference value.
Since all quantities of interest can be calculated using only configurational partition functions, these Monte Carlo calculations were mass-independent and considered only configurational ensemble averages.
In each case, a specialized ensemble average involving only the degrees of freedom ahead of the crack tip was utilized, which is exact and allows for greater efficiency.

\subsection{Isometric ensemble}\label{sec:monte-carlo.isometric}

The isometric ensemble configurational integral from Eq.~\eqref{eq:Q_con_rewrite} can be rewriten in terms of the Helmholtz free energy of the reference system using Eq.~\eqref{eq:AP} as

\begin{equation}
    Q_\mathrm{con}(v) =
    \int d\lambda \ e^{-\beta A_0(v,\boldsymbol{\lambda})} \, e^{-\beta U_1(\boldsymbol{\lambda})}
    .
\end{equation}
Scaling by the configuration integral at $v=1$ yields

\begin{equation}
    \begin{aligned}
        \frac{Q_\mathrm{con}(v)}{Q_\mathrm{con}(1)} =& \
        \frac{
            \int d\lambda \ e^{-\beta A_0(v,\boldsymbol{\lambda})} \, e^{-\beta U_1(\boldsymbol{\lambda})}
        }{
            \int d\lambda \ e^{-\beta A_0(1,\boldsymbol{\lambda})} \, e^{-\beta U_1(\boldsymbol{\lambda})}
        }
        \\ =& \
        \frac{
            \int d\lambda \ e^{-\beta\Delta A_0(v,\boldsymbol{\lambda})} \, e^{-\beta A_\star(\boldsymbol{\lambda})}
        }{
            \int d\lambda \ e^{-\beta A_\star(\boldsymbol{\lambda})}
        }
        ,
    \end{aligned}
\end{equation}
where $\Delta A_0(v,\boldsymbol{\lambda}) \equiv A_0(v,\boldsymbol{\lambda}) - A_0(1,\boldsymbol{\lambda})$ is the relative Helmholtz free energy of the reference system, and where $A_\star(\boldsymbol{\lambda}) \equiv A_0(1,\boldsymbol{\lambda}) + U_1(\boldsymbol{\lambda})$ is a convenient free energy.
Defining the specialized ensemble average

\begin{equation}
    \left\langle \phi\right\rangle_\star \equiv
    \frac{
        \int d\lambda \ e^{-\beta A_\star(\boldsymbol{\lambda})} \, \phi(\boldsymbol{\lambda})
    }{
        \int d\lambda \ e^{-\beta A_\star(\boldsymbol{\lambda})}
    }
    ,
\end{equation}
Eq.~\eqref{eq:betaDeltaAandp} is then used to write the relative nondimensional Helmholtz free energy of the full system as

\begin{equation}
    \beta\Delta A(v) =
    -\ln\left\langle e^{-\beta\Delta A_0}\right\rangle_\star
    .
\end{equation}
Computing $\partial/\partial v$ then yields the nondimensional force

\begin{equation}
    p(v) =
    e^{\beta\Delta A}\left\langle p_0 \, e^{-\beta\Delta A_0}\right\rangle_\star
    ,
\end{equation}
where $p_0\equiv\partial\beta\Delta A_0/\partial v$ is the nondimensional force of the reference system in the isometric ensemble.
Eq.~\eqref{eq:AP} is used to rewrite Eq.~\eqref{eq:k.isometric} as

\begin{equation}
    \frac{k'(v)}{k'(1)} =
    \frac{Q_\mathrm{con}^\ddagger(v)}{Q_\mathrm{con}^\ddagger(1)} \frac{Q_\mathrm{con}(1)}{Q_\mathrm{con}(v)} =
	e^{-\beta\Delta A^\ddagger(v)}e^{\beta\Delta A(v)}
    ,
\end{equation}
where $\Delta A^\ddagger(v) \equiv A^\ddagger(v) - A^\ddagger(1,\boldsymbol{\lambda})$ is the relative Helmholtz free energy of the full system with the crack tip bond fixed at its transition state.
The relative rate of breaking the crack tip bond is then

\begin{equation}
    \frac{k'(v)}{k'(1)} =
    \frac{\big\langle e^{-\beta\Delta A_0^\ddagger}\big\rangle_\star^\ddagger}{\big\langle e^{-\beta\Delta A_0}\big\rangle_\star}
    ,
\end{equation}
where the special transition state ensemble average $\langle\phi\rangle_\star^\ddagger$ uses $A_\star^\ddagger\equiv A_\star|_{\lambda_1=\lambda^\ddagger}$ and integrates over $\{\lambda_2,\ldots\lambda_M\}$.

\subsection{Isotensional ensemble}\label{sec:monte-carlo.isotensional}

The isotensional ensemble configurational integral from Eq.~\eqref{eq:Z_con_rewrite} can be rewriten in terms of the Gibbs free energy of the reference system using Eq.~\eqref{eq:GV} as

\begin{equation}
    Z_\mathrm{con}(p) =
    \int d\lambda \ e^{-\beta G_0(p,\boldsymbol{\lambda})} \, e^{-\beta U_1(\boldsymbol{\lambda})}
    .
\end{equation}
Scaling by the configuration integral at $p=0$ yields

\begin{equation}
    \begin{aligned}
        \frac{Z_\mathrm{con}(p)}{Z_\mathrm{con}(0)} =& \
        \frac{
            \int d\lambda \ e^{-\beta G_0(p,\boldsymbol{\lambda})} \, e^{-\beta U_1(\boldsymbol{\lambda})}
        }{
            \int d\lambda \ e^{-\beta G_0(0,\boldsymbol{\lambda})} \, e^{-\beta U_1(\boldsymbol{\lambda})}
        }
        \\ =& \
        \frac{
            \int d\lambda \ e^{-\beta\Delta G_0(p,\boldsymbol{\lambda})} \, e^{-\beta G_\star(\boldsymbol{\lambda})}
        }{
            \int d\lambda \ e^{-\beta G_\star(\boldsymbol{\lambda})}
        }
        ,
    \end{aligned}
\end{equation}
where $\Delta G_0(p,\boldsymbol{\lambda}) \equiv G_0(p,\boldsymbol{\lambda}) - G_0(0,\boldsymbol{\lambda})$ is the relative Gibbs free energy of the reference system, and where $G_\star(\boldsymbol{\lambda}) \equiv G_0(0,\boldsymbol{\lambda}) + U_1(\boldsymbol{\lambda})$ is a convenient free energy.
Defining the specialized ensemble average

\begin{equation}
    \left\langle \phi\right\rangle_\star \equiv
    \frac{
        \int d\lambda \ e^{-\beta G_\star(\boldsymbol{\lambda})} \, \phi(\boldsymbol{\lambda})
    }{
        \int d\lambda \ e^{-\beta G_\star(\boldsymbol{\lambda})}
    }
    ,
\end{equation}
Eq.~\eqref{eq:betaDeltaGandv} is then used to write the relative nondimensional Gibbs free energy of the full system as

\begin{equation}
    \beta\Delta G(p) =
    -\ln\left\langle e^{-\beta\Delta G_0}\right\rangle_\star
    .
\end{equation}
Computing $-\partial/\partial p$ then yields the nondimensional end separation

\begin{equation}
    v(p) =
    e^{\beta\Delta G}\left\langle v_0 \, e^{-\beta\Delta G_0}\right\rangle_\star
    ,
\end{equation}
where $v_0\equiv -\partial\beta\Delta G_0/\partial p$ is the nondimensional end separation of the reference system in the isotensional ensemble.
Eq.~\eqref{eq:GV} is used to rewrite Eq.~\eqref{eq:k.isotensional} as

\begin{equation}
    \frac{k'(p)}{k'(0)} =
    \frac{Z_\mathrm{con}^\ddagger(p)}{Z_\mathrm{con}^\ddagger(0)} \frac{Z_\mathrm{con}(0)}{Z_\mathrm{con}(p)} =
	e^{-\beta\Delta G^\ddagger(p)}e^{\beta\Delta G(p)}
    ,
\end{equation}
where $\Delta G^\ddagger(p) \equiv G^\ddagger(p) - G^\ddagger(0,\boldsymbol{\lambda})$ is the relative Gibbs free energy of the full system with the crack tip bond fixed at its transition state.
The relative rate of breaking the crack tip bond is then

\begin{equation}
    \frac{k'(p)}{k'(0)} =
    \frac{\big\langle e^{-\beta\Delta G_0^\ddagger}\big\rangle_\star^\ddagger}{\big\langle e^{-\beta\Delta G_0}\big\rangle_\star}
    ,
\end{equation}
where the special transition state ensemble average $\langle\phi\rangle_\star^\ddagger$ uses $G_\star^\ddagger\equiv G_\star|_{\lambda_1=\lambda^\ddagger}$ and integrates over $\{\lambda_2,\ldots\lambda_M\}$.

\bibliography{main}

\end{document}